\documentclass[preprint,12pt]{elsarticle}

\usepackage{amssymb}
\usepackage{amsmath}

\usepackage{lineno}
\usepackage{hyperref}
\hypersetup{
    colorlinks=true,
    linkcolor=blue,
    filecolor=magenta,      
    urlcolor=cyan,
    pdftitle={Overleaf Example},
    pdfpagemode=FullScreen,
    }

\newcommand{\ttbar}{$t\bar{t}$~}
\newcommand{\GeV}{~\text{GeV}}
\newcommand{\TeV}{~\text{TeV}}

\journal{Nuclear Physics B}

\begin{document}
\begin{frontmatter}



\title{Differential top quark cross section results from the ATLAS and CMS experiments}


\author[]{Johannes Hessler$^{~a,b}$  on behalf of the ATLAS and CMS Collaborations\footnote{Copyright 2026 CERN for the benefit of the ATLAS and CMS Collaborations. Reproduction of this article
or parts of it is allowed as specified in the CC-BY-4.0 license}} 
\affiliation[mpp]{organization={Max Planck Insitute for Physics},
            addressline={Boltzmannstr. 8}, 
            city={85748 Garching},
            country={Germany}}

\affiliation[tum]{organization={Technical University Munich},
            addressline={Arcisstraße 21}, 
            city={80333 Munich},
            country={Germany}}

\begin{abstract}
This report summarizes recent results of differential top quark cross section measurements performed by the ATLAS and CMS experiments.
The \ttbar process is studied as well as the production of single (anti-)top quarks and the interference with other Standard Model processes of the same final state.
State-of-the-art theory predictions are compared to the data.
No theory model is able to describe the data across all bins, but an improved description of the data when moving to predictions in higher orders in perturbative QCD can be observed.

\end{abstract}






\end{frontmatter}


\section{Introduction}
\label{sec:Introduction}
\noindent
The top quark is the heaviest known particle in the Standard Model (SM) of particle physics.
It shows the strongest coupling to the Higgs boson and is of great value for
precision tests of the SM, probing the limits of perturbative
Quantum Chromodynamics (QCD) at next-to-next-to-leading-order (NNLO) accuracy.
A rich set of physics results was obtained by the ATLAS~\cite{ATLAS_experiment} and CMS~\cite{CMS_experiment} experiments, analyzing the full Run 2 dataset taken at $\sqrt{s}=13 \TeV$ with an integrated luminosity of $140~\text{fb}^{-1}$ and $138~\text{fb}^{-1}$, respectively.

Top quark-antiquark ($t\bar{t}$) pairs are dominantly produced via gluon-gluon fusion in proton-proton collisions at the Large Hadron Collider at CERN.
A top quark almost exclusively decays into a $W$-boson and a $b$-quark. The events are generally categorized by the decay modes of the two $W$-bosons produced in a \ttbar event.
The all-hadronic channel is most abundant ($46\%$), followed by the single lepton ($45\%$) and the dilepton channel ($9\%$).
The same $WbWb$ final state as from \ttbar events, can also originate from processes with only one top quark or electro-weak processes without any intermediate tops. Two ATLAS analyses therefore follow a more inclusive approach, defining the final state directly as $WbWb$.

The differential cross sections are measured as functions of top quark and decay products kinematic properties. Dedicated analyses focus on top jet substructure, single top quark production and comparisons of the \ttbar process to $WbWb$ production.
The obtained results can be used to constrain Monte-Carlo (MC) generator parameters
in the top quark decay, the parton shower (PS) and the modeling of SM backgrounds involving top quarks.



\section{Top quark pair production in the single-lepton channel}
\noindent
Differential cross sections with exactly one electron or muon in addition to several jets are presented by the CMS collaboration~\cite{CMS_singlelepton_2021}. The extensive event sample collected during Run 2 ($\sqrt{s}=13 \TeV, L=138~\text{fb}^{-1}$) and the large branching fraction and small background contamination of the single lepton channel are exploited.
The reconstruction techniques for resolved \ttbar events $(p_T(t)<500 \GeV)$ introduced in previous analyses~\cite{CMS_singlelepton_2018} are extended by adding Lorentz-boosted top quarks with overlapping decay products. This constitutes the first analysis combining resolved and boosted topologies to determine the full differential cross section spectrum. 

The jet assignment for the top reconstruction in the resolved region is handled by maximizing the likelihood function imposing constraints on the intermediate resonances, specifically the mass of the $W$-boson and the top quark as well as the overall transverse momentum. 
Hadronic boosted tops are identified as large radius jets ($R=0.8$) with a $b$-tag, whereas leptonic tops require a high-momentum lepton overlapping with a $b$-tagged small-radius jet ($R=0.4$).
Neural networks (NNs) are trained to distinguish signal events from background events in the boosted regime. 

The differential cross sections are presented at particle and parton level, where the latter represents a \ttbar pair before the decay. They are obtained by performing a combined fit to a total of 18 orthogonal categories defined by the reconstruction methods, the lepton flavors, and the three years of data taking. The systematic uncertainties are dominated by the jet energy scale, integrated luminosity and \ttbar modeling.
Thanks to the four times larger data set with respect to previous measurements and the enhanced analysis techniques, the measurement range of the momentum and mass observables is significantly increased to several TeV.
A set of MC event generator predictions are compared to the data.
The particle level differential cross sections are found to be compatible
with the standard model predictions of the event generators
\textsc{Powheg} interfaced with the Pythia or Herwig Parton Shower and \textsc{MG5}\_a \textsc{MC@NLO+Pythia}.
Calculations obtained with \textsc{Matrix} in NNLO QCD provide a good description of the parton level cross sections while providing significantly reduced theoretical uncertainties.

\section{\texorpdfstring{\ttbar}{tt} production in the dilepton channel}
\noindent
Complementary to the single-lepton channel, the CMS collaboration also provides recent results in \ttbar events where both $W$-bosons decay into a charged lepton and a neutrino~\cite{CMS_dilepton_2025}.
The resulting dilepton channel has a relatively small branching fraction,
but also significantly lower background compared to other decay channels.
The precision of observables in the top quark sector based on the leptons is unmatched 
thanks to the excellent lepton energy resolution.
With respect to previous analyses, uncertainties could be reduced by almost a factor of two and the range of kinematic observables was extended.
A kinematic fit is performed to define the \ttbar system by explicitly reconstructing the neutrino 4-vectors. An alternative, looser selection is provided which does not rely on the top quark mass as input. Thus, preserving the sensitivity for future top mass extractions.

The cross sections are extracted by regularized matrix inversion with the \textsc{TUnfold} package.
It makes use of the response matrix to model resolution, acceptance and efficiency effects in a given phase space.
A great variety of single-, double- and triple-differential cross sections are provided as function of variables defined on parton or particle level.
The total uncertainty on the cross section of individual bins ranges between 2 and $20\%$ and is mostly dominated by the jet energy scale uncertainty.
In addition to the kinematic properties of single top (anti-)quarks and \ttbar pairs, an extensive list of observables was measured. That includes studies of the correlations between top and anti-top quark which are directly related to charge asymmetry and PDF-sensitive quantities.
Valuable input to future extractions of PDFs is provided by measuring triple-differential cross sections as a function of the full kinematics of the \ttbar system (transverse momentum, invariant mass and rapidity) for the first time.

Predictions of several NLO MC event generators  were compared
to the data. The predictions of these MC models generally fail to describe many of the measured cross sections in their full kinematic range.
When also taking theoretical uncertainties of the predictions into account
tensions are significantly reduced.
Beyond-NLO theoretical predictions show a description of the data of similar or improved quality for most observables compared to \textsc{Powheg+Pythia}. An exception are some of the kinematic distributions that are directly sensitive to higher-order QCD corrections.
The obtained results provide important input for future theoretical predictions as well as valuable input to future top mass and PDF extractions.

\section{Jet substructure observables in boosted \texorpdfstring{$t\bar{t}$}{tt} events}
\noindent
A precise understanding of the top quark decay process and the consecutive
parton-showering and hadronization effects is crucial for analyses targeting final states with top quarks.
The ATLAS collaboration investigated the substructure of large-radius top jets in boosted \ttbar events~\cite{ATLAS_substructure_2024}. The jets are reconstructed with the anti-$k_t$ algorithm, where the jet radius is set to $R=1.0$ and only charged constituents are considered. Events where both top quarks decay fully-hadronically are considered, as well as semi-leptonic events with exactly one charged lepton in the final state.
The transverse momentum of the top quark jets are required to exceed $350\GeV$.
The data are unfolded to particle level with the Iterative Bayesian Unfolding (IBU) method.
Eight substructure variables are defined and cross sections are presented singly and doubly differential. The selected observables aim to provide a complete picture of the jet substructure, while also considering their relevance in tagging algorithms.
It was found that observables sensitive to the broadness such as the Les Houche angularities are in good agreement with serval Monte-Carlo event generators simulating the top quark pair production in NLO QCD. However, the models are in tension with the data when considering observables probing the three-prong structure, such as the $N$-jettiness observables.

\section{Differential cross sections of \texorpdfstring{$WbWb$}{WbWb} production in the dilepton channel}
\noindent
A more inclusive approach to top cross section measurements is followed in the $WbWb$ measurement performed by the ATLAS collaboration~\cite{ATLAS_wbwb_2025}.
The signal process definition comprises both, $t\bar{t}$ production, as well as single top production in association with a $W$-boson ($tW$).
Possible contributions of di-boson production in association with a $b$-quark pair are negligible in the selected phase space and are not considered as part of the signal.

The goal of the analysis is a more detailed understanding of $t\bar{t} / tW$ interference effects. Two schemes are employed to model the interference, diagram removal (DR) and diagram subtraction (DS). The measured cross sections are compared to several MC predictions, including the 
Powheg$+$Pythia \textit{bb4$\ell$} predictions. It generates the $l^+\nu l^- \bar{\nu}b\bar{b}$ final state directly, avoiding the ambiguity of the DR/DS description.

The analysis is performed in the di-lepton channel.
For the event selection, exactly one electron and one muon of opposite charge are required. At least two jets must be present and identified as originating from a $b$ quark.
The measured detector level distributions are unfolded into the fiducial particle level phase space with the IBU method.
The uncertainties are reduced by a factor of two with respect to previous analyses,
allowing a better discrimination between the predictions.
The dominating uncertainties stem primarily from signal modeling, including those
related to the treatment of the $t\bar{t}/tW$ interference, as well as from statistical uncertainties in the tails of the distributions.

The $m^{bl}_\text{minimax}$ observable was found to maximize the sensitivity to the DR/DS modeling.  In order to suppress combinatorial background and reduce impacts of mis-modeling at high $b$-jet multiplicities, exactly two $b$-tagged jets are required for the measurement of this observable. 
For doubly-resonant $t\bar{t}$ events, the observable is constrained by the top quark mass. This leads to an enhanced sensitivity to interference effects for the high-mass region above the top quark mass.

The low mass region is well described by all tested MC models. They agree with the data within the uncertainties. However, none of the predictions are able to describe the data up to the extreme tails of the mass distribution. While the predictions in the DR prescription show the best description of the data, they overestimate the data at the highest masses. The DS scheme on the other hand predicts a too low cross section in that region. The $bb4\ell$ model performs better, but also underestimates the high-mass tail.

The results of the analysis provide valuable insights into the modeling of the 
$WbWb$ final state. They highlight the need for further improvements in the simulation of the full $WbWb$ final state.
This would allow a more complete description of the interference and lead to reduced corresponding uncertainties.

\section{Single top (anti-)quark production in the \textit{t}-channel}
\noindent
The main production channel of single (anti-)top quarks at the LHC is the exchange of a virtual $W$-boson in the $t$-channel.
The $tq$ cross section is expected to be larger than the $\bar{t}q$ cross section, 
since the valence up quark density of the proton is about twice as high as the valence down quark density.
Sub-leading diagrams that are strictly sea quark induced contribute in a charge-symmetric way. Measured cross sections of single top quark production are sensitive to PDFs of the proton.

This analysis~\cite{ATLAS_singletop_2025}
performed by the ATLAS collaboration measures the differential production cross sections of single top and anti-top quarks.
The observables of interest are the transverse momentum and absolute rapidity of the quark measured separately for top and anti-top quarks.
In addition, the ratio of top to anti-top cross sections is measured.
The publication also comprises an interpretation of the results in an effective field theory approach which will not be discussed in the scope of this proceeding.

The event selection follows the procedure introduced for the inclusive single-top quark cross section measurement. 
Exactly one isolated electron or muon and a substantial amount of missing transverse energy are required, originating from the leptonic decay of the $W$-boson. A jet multiplicity of exactly two is enforced, based on the LO event signature.
One of the jets has to fulfill $b$-jet tagging requirements.
In addition to kinematic requirements on the final state objects, a NN is trained to separate between signal and background events.
Two signal regions for the $tq$ and $\bar{t}q$ cross section measurements are constructed, defined by charge of the final state lepton. \\
In order to correct the data for acceptance, resolution and efficiency effects, they are unfolded to parton level with the IBU method. They are extrapolated to the full kinematic phase space and can be directly compared to theory predictions.
The most relevant background contributions originate from other processes containing top quarks, such as \ttbar production and single top production in the $s$-channel or in association with a $W$-boson. 

Several predictions are compared to the data, including fixed order calculations (in LO, NLO and NNLO) and various PDF sets. 
Overall, a good agreement between the predictions and the data can be observed within the measurement uncertainties.
The dominating systematic uncertainties come from signal modeling. The statistical uncertainties become significant for normalized cross sections, as well as for cross section ratios.
The measured single top quark cross sections can put constraints of the parton distribution functions (PDFs) of the proton. The cross section ratio $tq/\bar{t}q$ as a function of $|y(t)|$ is of particular interest.

\section{Lepton differential distributions in \texorpdfstring{\ttbar}{tt} dilepton events}
\noindent
Differential cross sections for the process \ttbar$\rightarrow e\mu\nu\bar{\nu}b\bar{b}$ as well as complementary measurements of the inclusive $e\mu b\bar{b}$ production are provided by the ATLAS collaboration~\cite{ATLAS_dilepton_2025}. The latter includes \ttbar and $tW$ processes in the signal definition. The main results of the analysis, the precise measurement of the inclusive production cross section and the determination of the top quark mass from the inclusive cross section will not be discussed here.

Exactly one electron and one muon is required in the event selection. Differential cross section results on particle level are obtained by correcting the measured distributions with multiplicative bin-by-bin factors for smearing and acceptance effects induced by the detector.
The dominating background contribution for the \ttbar measurement consists of $tW$ events with two prompt leptons in the final state. The subtraction of these events from the signal events leads to one of the leading uncertainties on the final cross sections. Additional $b$-jet requirements are imposed on the event selection for the $e\mu b\bar{b}$ phase space. Normalization effects of the samples mostly cancel in the unfolding, leading to an overall reduced systematic uncertainty.

Single and double differential cross sections are reported as functions of single lepton quantities as well as of the combined lepton pair.
Both sets of cross sections are compared to various MC event generator predictions.
It can be observed, that the state-of-the-art generators
\textsc{Powheg} MiNNLO and \textsc{Powheg} $bb4\ell$ provide an improved description of the data over the \textsc{Powheg} \texttt{hvq} model which is traditionally used for LHC physics analyses.

\section{Summary}
\noindent
Top quarks are produced in abundance in proton-proton collisions at the LHC.
A wide range of top quark physics results is covered by the ATLAS and CMS collaborations.
This includes differential cross sections in single-top quark production, \ttbar production in the single- and dilepton final states, as well as the first direct measurement of the $WbWb$ final state.

Thanks to improved analysis techniques, the uncertainties could be reduced significantly with respect to previous analyses.
Theory predictions at NNLO QCD provide an improved description of the data.
However, none of the generators describe the data in all phase space regions.
Discrepancies can be observed in extreme regions of the phase space and in higher-dimensional distributions.

All presented results were obtained from LHC Run 2 data.
Many exciting results can be expected from Run 3, benefiting from the increased size of the data set, as well as 
refined reconstruction, calibration and analysis techniques.






\end{document}